\documentclass[seceq,preprint]{ptptex}

\usepackage{graphicx}
\usepackage{wrapft}

\newcommand{\beq}{\begin{equation}}
\newcommand{\eeq}{\end{equation}}
\newcommand{\ber}{\begin{eqnarray}}
\newcommand{\eer}{\end{eqnarray}}

\newcommand{\cD}{\mathcal{D}}

\title{%        %You can use \\ for explicit line-break
Nonperturbative QCD, gauge-fixing, Gribov copies, and the lattice}

\author{%       %Use \scshape  for the family name
Anthony G. \textsc{WILLIAMS}%
}

\inst{%         %Affiliation, neglected when [addenda] or [errata]
Special Research Centre for the Subatomic Structure
of Matter (CSSM), \\
University of Adelaide, SA 5005, Australia
}

\preprintnumber{ADP-03-118/T556}

\abst{%         %this abstract is neglected when [addenda] or [errata]
Perturbative QCD uses the Faddeev-Popov gauge-fixing procedure,
which leads to ghosts and the local BRST invariance of the
gauge-fixed perturbative QCD action.  In the asymptotic
regime, where perturbative QCD is relevant, Gribov copies
can be neglected.  In the nonperturbative
regime, one must adopt either a nonlocal Gribov-copy
free gauge (e.g., Laplacian gauge) or attempt to maintain
local BRST invariance at the expense of admitting Gribov copies.
These issues are explored.  In addition, we discuss the
relationship between recent Dyson-Schwinger based model calculations
of the infrared behavior of QCD Green's functions and the
lattice calculation of these quantities.
}

\begin{document}

\maketitle

\section{Introduction}

Perturbative quantum chromodynamics (QCD) is formulated using
the Faddeev-Popov gauge-fixing procedure, which introduces ghost fields
and leads to the local BRST invariance of the gauge-fixed perturbative
QCD action.  These perturbative gauge fixing schemes include, e.g., the
standard choices of covariant, Coulomb and axial gauge fixing.
These are entirely adequate for the purpose of
studying perturbative QCD, however, they fail in the
nonperturbative regime due to the presence of Gribov copies.
%i.e., a discrete set of gauge-equivalent gauge-field configurations survive
%in the gauge functional integration after these (incomplete) gauge fixings.
Perturbative QCD works because in doing a weak-field expansion
around the $A_\mu=0$ configuration these Gribov copies are not
encountered~\cite{AGW_summary}.

One could define nonperturbative
QCD by imposing a non-local Gribov-copy free gauge fixing
(such as Laplacian gauge) or, alternatively, one could
attempt to maintain local BRST invariance at the cost of admitting
Gribov copies.  One of the well-known difficulties for the latter
option is the problem of pairs of Gribov copies
with opposite sign giving a vanishing path 
integral\cite{Giusti:2001xf,vanBaal:1997gu,Neuberger:xz,Testa:1998az}.
Whether or not a local BRST invariance for QCD can be maintained in the
nonperturbative regime remains an open problem.

The standard lattice definition of QCD is equivalent
to the choice of a Gribov copy free gauge-fixing.  There is a 
negligible chance of selecting two gauge-equivalent configurations
(strictly zero except for numerical round-off error).  Calculations
of {\em physical observables} are unaffected by arbitrary gauge
transformations on the configurations in the ideal gauge-fixed
ensemble. A lattice QCD calculation using an ideal gauge-fixed
ensemble will give a result for a gauge-invariant (i.e., physical)
quantity which is identical to doing no gauge fixing at all, i.e.,
equivalent to the standard lattice calculation of physical quantities.

We begin by reviewing the standard arguments for constructing
QCD perturbation theory, which use the Faddeev-Popov gauge fixing
procedure to construct the perturbative QCD gauge-fixed Lagrangian
density.  The naive Lagrangian density of QCD is
$
{\mathcal{L}}_{\mathrm{QCD}} = -\frac{1}{4}F^{\mu\nu}F_{\mu\nu} 
                          + \sum_f{\bar{q}}_f (iD\!\!\!\!/-m_f)q_f,
$
where the index $f$ corresponds to the quark flavours. The naive
Lagrangian is neither gauge-fixed nor renormalized, however it is
invariant under local $SU(3)_c$ gauge  transformations $g(x)$.
For arbitrary, small $\omega^a(x)$ we have \\
$g(x) \equiv \exp\left\{-ig_s\left(\lambda^a/2\right)\omega^a(x)
                \right\} \in SU(3),
$
where the $\lambda^a/2\equiv t^a$ are the generators of the gauge group
$SU(3)$ and the index $a$ runs over the eight generator labels
$a=1,2,...,8$.

Consider some gauge-invariant Green's function
(for the time being we shall
concern ourselves only with gluons)
$
\langle \Omega|\ T(\hat O[A])\ |\Omega\rangle 
 = \int{\cD}A\ O[A]\ e^{iS[A]}/\int{\cD}A\ e^{iS[A]} \, ,
$
where $O[A]$ is some gauge-independent quantity depending on the gauge
field, $A_\mu(x)$.
We see that the gauge-independence of $O[A]$ and $S[A]$ 
gives rise to an infinite quantity in both
the numerator and denominator,
which must be eliminated by gauge-fixing.  The Minkowski-space
Green's functions are defined as the Wick-rotated versions of the
Euclidean ones.

\begin{figure}
\centerline{\includegraphics[height=3 cm]{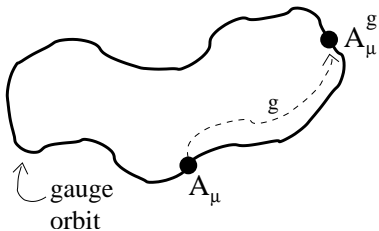}}
\caption{Illustration of the gauge orbit containing $A_\mu$ and indicating
the effect of acting on $A_\mu$ with the gauge transformation $g$.  The
action $S[A]$ is constant around the orbit.}
\label{fig:Gorbit}
\end{figure}

The gauge orbit for some configuration $A_\mu$ is defined to be the set of
all of its gauge-equivalent configurations. Each point $A_\mu^g$ on the gauge
orbit is obtained by acting upon $A_\mu$ with the gauge transformation $g$.
By definition the action, $S[A]$, is gauge invariant and so all
configurations on the gauge orbit have the same action, e.g., see the
illustration in Fig.~\ref{fig:Gorbit}.

\section{Gribov Copies and the Faddeev-Popov Determinant}

\begin{figure}
\centerline{\includegraphics[angle=0,width=4cm]{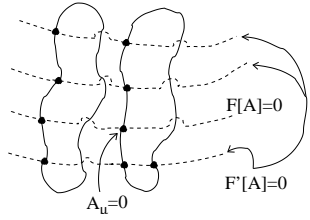}}
\caption{Ideal, $F[A]$, and non-ideal, $F'[A]$, gauge-fixing.}
\label{fig:manyorbit}
\end{figure}

Any gauge-fixing procedure defines a surface in gauge-field
configuration space. Fig.~\ref{fig:manyorbit} is a depiction of these
surfaces represented as dashed lines intersecting the gauge orbits within 
this configuration space. Of course, in general, the gauge orbits are
hypersurfaces as are the gauge-fixing surfaces.  Any gauge-fixing
surface must, by definition, only intersect the gauge orbits at distinct
isolated points in field configuration space.  For this reason, it is
sufficient to use lines for the simple illustration of the concepts here.
An ideal (or complete) gauge-fixing condition, $F[A]=0$, defines a surface
called the Fundamental Modular Region (FMR) that intersects each gauge
orbit once and only once and typically where possible
contains the trivial configuration $A_\mu=0$.  A non-ideal gauge-fixing
condition, $F'[A]=0$, defines a surface or surfaces which intersect the
gauge orbit more than once. These multiple intersections of the non-ideal
gauge fixing surface(s) with the gauge orbit are referred to as Gribov
copies
\cite{Giusti:2001xf,vanBaal:1997gu,Neuberger:xz,Testa:1998az}.
Lorentz gauge ($\partial_\mu A^\mu(x)=0$) for example, has many
Gribov copies per gauge orbit.  By definition an ideal gauge
fixing is free from Gribov copies.  The ideal gauge-fixing
surface $F[A]=0$ specifies the FMR for that gauge
choice.  Typically the gauge fixing condition depends on a space-time
coordinate, (e.g., Lorentz gauge, axial gauge, {\em etc.}), and so we
write the gauge fixing condition more generally as $F([A];x)=0$. 

Let us denote one arbitrary gauge configuration per gauge orbit,
$A_\mu^0$, as the origin for that gauge orbit, i.e., corresponding to
$g=0$ on that orbit.  Then each gauge orbit can be labelled by
$A_\mu^0$ and the set of all such $A_\mu^0$ is equivalent to one
particular, complete specification of the gauge.
Under a gauge transformation, $g$, we move from the origin of the gauge
orbit to the configuration, $A_\mu^g$, where by definition
$A_\mu^0 \stackrel{g}{\longrightarrow} A_\mu^g = 
    g A_\mu^0 g^\dagger - (i/g_s)(\partial_\mu g)g^\dagger$.
Let us denote for each gauge orbit the gauge transformation,
$\tilde g\equiv \tilde g[A^0]$, as the transformation which
takes us from the origin of that
orbit, $A_\mu^0$, to the corresponding configuration on the FMR, 
$A_\mu^{\rm FMR} \equiv A_\mu^{\tilde g}$,
which is specified by the ideal gauge fixing condition
$F([A^g];x)=0$.
In other words, an ideal gauge fixing has a unique $\tilde g$ which
satisfies $ F([A^g];x)|_{\tilde g}=0$ and hence specifies the FMR as 
$A^{\tilde g}\equiv A_\mu^{\rm FMR}\in {\mathrm {FMR}}$.  Note then
that we have
$  \int{\cD}A = \int{\cD}A^0\int{\cD}g = 
 \int{\cD}A^{\rm FMR}\int{\cD}(g-\tilde g) \, .$

The {\em inverse Faddeev-Popov determinant} is defined as the integral over
the gauge group of the gauge-fixing condition, i.e.,
\ber
\Delta^{-1}_F[A^{\rm FMR}] \equiv \int{\cD}g\ \delta[F[A]]
   = \int{\cD}g\ \delta(g-{\tilde{g}}) 
   \left|\det\left(\frac{\delta F([A];x)}{\delta g(y)}\right)
                              \right|^{-1} 
\eer
Let us define the matrix $M_F[A]$ as
$
M_F([A];x,y)^{ab}\equiv \delta F^a([A];x)/\delta g^b(y) \, .
$
Then the {\em Faddeev-Popov determinant} for an arbitrary configuration
$A_\mu$ can be defined as
$\Delta_F[A] \equiv \left|\det M_F[A]\right|$.
(The reason for the name is now clear).  Note that we
have consistency, since
$\Delta^{-1}_F[A^{\rm FMR}] \equiv \Delta^{-1}_F[A^{\tilde g}]
= \int{\cD}g\ \delta(g-{\tilde{g}})\Delta^{-1}_F[A]$.

We have $1 = \int{\cD}g\ \Delta_F[A]\,\, \delta[F[A]] =
\int{\cD}(g-\tilde g)\ \Delta_F[A]\,\, \delta[F[A]]$
by definition and hence
\ber
  \int\!{\cD}A^{\rm FMR}\! \equiv\!\!  
          \int\!{\cD}A^{\rm FMR}\!\!\!\int\!{\cD}(g-\tilde g)\ 
                \Delta_F[A]\delta[F[A]]
             \! = \!\! \int\!{\cD}A\ \Delta_F[A] \delta[F[A]]
\eer
Since for an ideal gauge-fixing there is one and only one $\tilde{g}$
per gauge orbit, such that
$F([A];x)|_{\tilde g}=0$, then $|$det$M_F[A]|$ is non-zero on the FMR. 
It follows that since there is at least one smooth path between any two
configurations in the FMR and since the determinant cannot be zero on the 
FMR, then it cannot change sign on the FMR. The 
{\em first Gribov horizon} is defined to be those configurations with
$\det M_F[A]=0$ which lie closest to the FMR.  By
definition the determinant can change sign on or outside this horizon.
Clearly,
the FMR is contained within the first Gribov horizon and for an ideal gauge
fixing, since the sign of the determinant cannot change, we can replace
$|\det M_F|$ with $\det M_F$, [i.e., the overall sign of the functional
integral is normalized away in the ratio of functional integrals].

These results are generalizations of results from ordinary calculus,
where \\
$
\left|{\mathrm {det}}\left(\partial f_i/\partial x_j\right)
        \right|_{\vec{f}=0}^{-1}
           = \int dx_1\cdots dx_n\, \delta^{(n)}({\vec{f}}({\vec{x}}))
$ \\
and if there is one and only one ${\vec{x}}$ which is a solution of
${\vec{f}}({\vec{x}})=0$ then the matrix
$M_{ij} \equiv \partial f_i/\partial x_j$
is invertible (i.e., non-singular) on the hypersurface 
${\vec{f}}({\vec{x}})=0$ and hence $\det M\neq 0$. 

\section{Generalized Faddeev-Popov Technique}

Let us now assume that we have a family of {\em ideal}
gauge fixings $F([A];x)=f([A];x)-c(x)$
for any Lorentz scalar $c(x)$ and for $f([A];x)$ being some Lorentz scalar
function,
(e.g., $\partial^\mu A_\mu(x)$ or $n^\mu A_\mu(x)$ or similar or any
nonlocal generalizations of these).
Therefore, using the fact that we remain in the FMR and can drop the
modulus on the determinant, we have
$
\int{\cD}A^{\rm FMR} = \int{\cD}A\ \det M_F[A]\ \delta[f[A]-c]
     \, .
$
Since $c(x)$ is an arbitrary function, we can define a new
``gauge'' as the Gaussian weighted average over $c(x)$, i.e.,
\begin{eqnarray}
\int{\cD}A^{\rm FMR} &\propto& \int{\cD}c\ {\mathrm{exp}}\left\{
                      -\frac{i}{2\xi}\int d^4x c(x)^2 \right\}
                      \int{\cD}A\ {\mathrm {det}} M_F[A]\
                                   \delta[f[A]-c] \nonumber \\
                  & \propto & \int{\cD}A\ {\mathrm {det}}M_F[A]
                      {\mathrm{exp}}\left\{-\frac{i}{2\xi}
                     \int d^4x f([A];x)^2 \right\} \nonumber \\
                  & \propto & \int{\cD}A {\cD}\chi{\cD}{\bar{\chi}}\ \ 
                      {\mathrm{exp}}\left\{
                      -i\int d^4xd^4y \frac{}{} 
                     {\bar{\chi}}(x)M_F([A];x,y)\chi(y)\right\} \nonumber \\
                  & & \times {\mathrm{exp}}\left\{
                      -\frac{i}{2\xi}\int d^4x f([A];x)^2 \right\} \, ,
\end{eqnarray}
where we have introduced the anti-commuting ghost fields
$\chi$ and $\bar{\chi}$.
Note that this kind of ideal gauge fixing does not choose just one gauge
configuration on the gauge orbit, but rather is some Gaussian
weighted average over gauge fields on the gauge orbit.
We then obtain
\ber
\langle\Omega|\ T(\hat{O}[...])\ |\Omega\rangle 
       = \frac{\int{\cD}q{\cD}\bar{q}{\cD}A{\cD}\chi{\cD}{\bar{\chi}}\
        O[...]\ e^{iS_{\xi}[...]}}
         {\int{\cD}q{\cD}\bar{q}{\cD}A{\cD}\chi{\cD}{\bar{\chi}}\ 
        e^{iS_{\xi}[...]}}\, , 
\eer
where 
\begin{eqnarray}
S_{\xi}[q,\bar{q},A,\chi,{\bar{\chi}}]\! & = &\!\! \int\! d^4x\! \left[
   -\frac{1}{4}F^{a\mu\nu}F^a_{\mu\nu} -\frac{1}{2\xi}\left(f([A];x)\right)^2  
      \! +\! \sum_f{\bar{q}}_f (iD\!\!\!\!/-m_f)q_f\right]  \nonumber \\
   & & + \int d^4xd^4y\ {\bar{\chi}}(x)M_F([A];x,y)\chi(y) \, .
\label{Eq:gen_action}
\end{eqnarray} 

\section{Standard Gauge Fixing}

  We can now recover standard gauge fixing schemes as special cases
of this generalized form.  First consider standard covariant
gauge, which we obtain by taking
$f([A];x)=\partial_\mu A^{\mu}(x)$ and by {\em neglecting} the fact that this
leads to Gribov copies.
We need to evaluate $M_F[A]$ in the vicinity of the gauge-fixing
surface (specified by $\tilde g$):
\begin{eqnarray}
M_F([A];x,y)^{ab} = \frac{\delta F^a([A];x)}{\delta g^b(y)}
        = \frac{\delta [\partial_\mu A^{a\mu}(x)-c(x)]}{\delta g^b(y)} 
        = \partial^x_\mu\frac{\delta A^{a\mu}(x)}{\delta g^b(y)]} \, .
\end{eqnarray}
Under an infinitesimal gauge transformation about the FMR,
$\delta g\equiv g-\tilde g$, we have 
$(A^{\tilde g})_\mu\to (A^{\tilde g + \delta g})_\mu$, 
where 
\ber
  (A^{\tilde g+\delta g})_\mu^a(x) = (A^{\tilde g})_\mu^a(x) 
                + g_sf^{abc}\omega^b(x)A_\mu^c(x)
                -\partial_\mu\omega^a(x) 
                + {\mathcal O}(\omega^2)
\eer
and hence near the gauge fixing surface (i.e., for
small fluctations along the orbit around $A_\mu^{\rm FMR}$) using
$M_F([A];x,y)^{ab} \equiv \partial^x_\mu [\delta A^{a\mu}(x)/
\delta (\delta\omega^b(y)])|_{\omega=0}$ we find
\begin{eqnarray}
M_F([A];x,y)^{ab} 
%%%   \hskip-0.1cm
%%%   = \hskip-0.1cm \left. \partial^x_\mu
%%%   \frac{\delta A^{a\mu}(x)}{\delta (\delta\omega^b(y)]} \right|_{\omega=0}
%%%   \hskip-3.2cm 
        = \partial^x_\mu \left(\frac{}{}[-\partial^{x\mu}\delta^{ab} 
               + g_s f^{abc}A^{c\mu}(x)] 
%%%                    \right.  \nonumber \\ & & \left. 
                \delta^{(4)}(x-y)\right) \, . \nonumber 
\end{eqnarray}                                 
We then recover the standard covariant gauge-fixed form of the QCD action
\begin{eqnarray}
S_{\xi}[q,\bar{q},A,\chi,{\bar{\chi}}]
  &=& \int d^4x \left[
   -\frac{1}{4}F^{a\mu\nu}F^a_{\mu\nu} 
   -\frac{1}{2\xi}\left(\partial_\mu A^{\mu}\right)^2 
         + \sum_f{\bar{q}}_f (iD\!\!\!\!/-m_f)q_f\right] \nonumber \\
   && + (\partial_\mu\bar{\chi}_a)(\partial^\mu\delta^{ab}
       -gf_{abc}A_c^\mu)\chi_b \, .
\end{eqnarray}
However, this gauge fixing has not removed the Gribov copies and so
the formal manipulations which lead to this action are not valid. 
This Lorentz covariant set of naive gauges corresponds to a Gaussian
weighted average over generalized Lorentz gauges, where the gauge parameter
$\xi$ is the width of the Gaussian distribution over the configurations
on the gauge orbit.
Setting $\xi=0$ we see that the width vanishes and we obtain Landau 
gauge (equivalent to Lorentz gauge, $\partial^\mu A_\mu(x)=0$).
Choosing $\xi=1$ is referred to as ``Feynman gauge'' and so on.
We can similarly derive the QCD action for axial gauge.

 We can similarly recover the standard QCD action for the axial
 gauges, where 
 $n_\mu A^\mu(x)=0$.
 Proceeding as for the generalized covariant gauge, we first identify
 $f([A];x)=n_\mu A^\mu(x)$ and obtain the gauge-fixed action
 \begin{eqnarray}
 \hskip-0.2cm S_{\xi}[q,\bar{q},A] = \int d^4x 
      \left[-\frac{1}{4}F^{a\mu\nu}F^a_{\mu\nu}
        -\frac{1}{2\xi}\left(n_\mu A^\mu\right)^2
        + \sum_f{\bar{q}}_f (iD\!\!\!\!/-m_f)q_f\right] \, .
 \end{eqnarray}
 Taking the ``Landau-like'' zero-width limit $\xi \rightarrow 0$ we
 select $n_\mu A^\mu(x)=0$ exactly and recover the usual axial-gauge
 fixed QCD action.  Axial gauge does not involve ghost fields, since
 in this case
 \ber
 M_F([A^{\rm FMR}];x,y)^{ab}
     = \left. n_\mu\frac{\delta A^{a\mu}(x)}
          {\delta \omega^b(y)]} \right|_{\omega=0}
       =\! n_\mu
       \left([-\partial^{x\mu}\delta^{ab}]\delta^{(4)}(x-y)\right)
       \! ,
 \eer
 which is independent of $A_\mu$ since $n^\mu A_\mu^{\rm FMR}(x)=0$. 
 In other words, the gauge field does not appear in $M_F[A]$ on the
 gauge-fixed surface.
 Unfortunately axial gauge suffers from singularities which lead to
 significant difficulties when trying to define perturbation theory
 beyond one loop. A related feature is that axial gauge is
 not a complete gauge fixing prescription.  While there are complete
 versions of axial gauge on the periodic lattice, these always involve
 a nonlocal element, or reintroduce Gribov copies at the boundary so as
 not to destroy the Polyakov loops in the axial gauge-fixing direction.

\section{Discussion and Conclusions}

There is no known Gribov-copy-free gauge fixing which is a {\em local}
function of $A_{\mu}(x)$.  In other words, such a gauge fixing cannot
be expressed as a function of $A_\mu(x)$ and a finite number of its
derivatives, i.e., $F([A];x)\neq F(\partial_\mu,A_\mu(x))$ for all $x$.
Hence, the ideal gauge-fixed action, $S_{\xi}[\cdots]$, in
Eq.~(\ref{Eq:gen_action})
becomes non-local and gives rise to a nonlocal quantum field theory.
Since this action serves as the basis for the proof of the renormalizability
of QCD, the proof of asymptotic freedom, local BRST invariance, and the
Dyson-Schwinger equations\cite{DSE_review,Alkofer:2000wg}
(to name but a few) the nonlocality of the action
leaves us without a first-principles proof of these features of
QCD in the nonperturbative context.

The lattice implementation of Landau gauge finds local minima of the
gauge fixing functional, which correspond to configurations lying
inside the first Gribov horizon.  The remaining Gribov copies
after this partial gauge fixing then necessarily all have the
same sign (positive) for the Faddeev-Popov determinant and hence add
coherently in the functional integral.  This ensures that the
ghost propagator is positive 
definite\cite{Alkofer:2000wg,Zwanziger}.
The derivation of the Dyson-Schwinger equations is based on the fact
that the integral of a total derivative vanishes\cite{DSE_review}
provided that the surface integral of the integrand vanishes when
integrated over the boundary of the region.
Since the Faddeev-Popov determinant vanishes on
the first Gribov horizon, then we can still derive the standard
Dyson-Schwinger equations from the Landau gauge fixed QCD action even
if we restrict the gauge fields to lie within the first Gribov
horizon.  This is equivalent to requiring that the ghost propagator
be positive definite.  Thus it is valid to compare lattice
Landau-gauge calculations with Dyson-Schwinger based
calculations (with a positive definite ghost propagator), since these 
both consist of considering configurations within the first Gribov horizon.
An extensive body of lattice calculations exist for the Landau gauge
gluon\cite{gluon_Landau} and quark\cite{quark,quark_overlap}
propagators and most recently for the quark-gluon 
vertex\cite{quark_gluon_vertex}.  Similarly, calculations in Laplacian
gauge (an ideal gauge) fixing have also recently become 
available\cite{gluon_Laplacian,quark_asqtad}.

It is well-established that QCD is asymptotically free, i.e., it is
weak-coupling at large momenta.  In the weak coupling limit the functional
integral is dominated by small action configurations. As a consequence,
momentum-space Green's functions at large momenta will receive their
dominant contributions in the path integral from configurations near the
trivial gauge orbit, i.e., the orbit containing $A_\mu=0$, since this
orbit minimizes the action.  If we use standard
lattice gauge fixing, which neglects the fact that Gribov copies are
present, then
at large momenta $\int{\cD}A$ will be dominated by configurations lying
on the gauge-fixed surfaces in the neighbourhood of {\em each} of the Gribov
copies on the trivial orbit.  Since for small field fluctuations the Gribov
copies cannot be aware of each other, we merely overcount the contribution
by a factor equal to the number of copies on the trivial orbit.  This
overcounting is normalized away in the ratio of functional integrals.
Thus it is possible to understand why Gribov
copies can be neglected at large momenta and why it is sufficient to
use standard gauge fixing schemes as the basis for calculations in
perturbative QCD.

%\section*{Acknowledgements}
%We would like to thank ...........

%\appendix
%\section{First Appendix} %Empty argument \section{} yields `Appendix'. 
%
%\section{Second Appendix}

\end{document}